\documentstyle[12pt]{article}
\input epsf
\newcommand{\NPB}[3]{\emph{ Nucl.~Phys.} \textbf{B#1} (19#2) #3}   
\newcommand{\PLB}[3]{\emph{ Phys.~Lett.} \textbf{B#1} (19#2) #3}   
\newcommand{\PRD}[3]{\emph{ Phys.~Rev.} \textbf{D#1} (19#2) #3}

\def\dalemb#1#2{{\vbox{\hrule height .#2pt
        \hbox{\vrule width.#2pt height#1pt \kern#1pt
                \vrule width.#2pt}
        \hrule height.#2pt}}}

 \def\bd{\begin{document}} \def\ed{\end{document}}
\def\ds{\documentstyle} \let\fr=\frac \let\bl=\bigl \let\br=\bigr
\let\Br=\Bigr \let\Bl=\Bigl 
\let\bm=\bibitem
\let\na=\nabla
\let\pa=\partial \let\ov=\overline
\def\ie{{\it i.e.\ }} 
\def\beq{\begin{equation}}
\def\eeq{\end{equation}}
\def\beqa{\begin{eqnarray}}
\def\eeqa{\end{eqnarray}}

\newcommand{\ba}{\begin{array}}
\newcommand{\ea}{\end{array}}
\newcommand{\td}{\tilde}
\newcommand{\norsl}{\normalsize\sl}
\newcommand{\ns}{\normalsize}
\newcommand{\refs}[1]{(\ref{#1})}
\def\simlt{\mathrel{\lower2.5pt\vbox{\lineskip=0pt\baselineskip=0pt
           \hbox{$<$}\hbox{$\sim$}}}}
\def\simgt{\mathrel{\lower2.5pt\vbox{\lineskip=0pt\baselineskip=0pt
           \hbox{$>$}\hbox{$\sim$}}}}

\begin{document}
\thispagestyle{empty}
\rightline{CPHT-S-751.1299}
\rightline{CTP-TAMU-47-99}
\rightline{CERN-TH/99-385}
\rightline{\large\sf hep-ph/9912287}
\vskip 1.0truecm
\centerline{\Large\bf Looking For TeV-Scale Strings and
Extra-Dimensions}
\vskip 1.truecm
\centerline{{\large\bf E. Accomando}~$^a$, {\large\bf I. Antoniadis}~$^b$ 
and {\large\bf K. Benakli~$^c$}}
\vskip .5truecm
\centerline{{\it $^a$Center for Theoretical Physics, Texas A\&M University,
College Station, TX77843-4242, USA }}
\centerline{\it $^b$ Centre de Physique Th\'eorique, Ecole
Polytechnique, 91128 Palaiseau, France}
\centerline{\it Unit\'e mixte du CNRS et de l'EP, UMR 7644}
\centerline{{\it $^c$CERN Theory Division
 CH-1211, Gen\`eve 23, Switzerland}}
\vskip .5truecm

\vskip 1.truecm
\centerline{\bf\small ABSTRACT}
\vskip .5truecm

In contrast to the old heterotic string case, the (weakly coupled) type I
brane framework allows to have all, part or none of the standard model gauge
group factors propagating in large extra--dimensions of TeV$^{-1}$ size. 
We investigate the main experimental signatures of these possibilities,
related to the production of Kaluza-Klein excitations of gluons and
electroweak gauge bosons. A discovery through direct observation of
resonances is  possible only for compactification scales below 6 TeV. 
However effects due to exchange of virtual Kaluza-Klein excitations
could be observed for higher scales. We find that LHC can probe
compactification scales as high as 20 TeV for excitations of gluons and
8-15 TeV for excitations of electroweak gauge bosons.
Finally, in the case where no gauge boson feels the extra-dimension, we find
that effective contact interactions due to massive string mode oscillations
dominate those due to the exchange of Kaluza-Klein excitations of gravitons
and could be used to obtain bounds on the string scale.

\vskip 1.5truecm

\begin{flushleft}
\today
\end{flushleft}

\vfill\eject

\section{Introduction}

A lot of efforts have been devoted to understand possible patterns of new
physics beyond the standard model. One of the most spectacular possibilities is
the proposal of existence of large extra--dimensions compactified in the TeV
range \cite{Ant}. Such a scenario is easily realized if the fundamental string
or quantum gravity scale is low \cite{add}-\cite{st}.

Previous studies of the phenomenological implications of such 
extra--dimensions were focused on the simplest scenario where all standard model
gauge bosons propagate in the same compact space \cite{IK}-\cite{Nath}.
Although this is the simplest scenario favored by gauge coupling unification
\cite{unif,unif2}, it does not constitute the most general case. The different factors
of the standard model gauge group may arise from branes of different kinds,
extended in different compact directions. As a result, TeV-dimensions can be
longitudinal along the world-volume of some branes and transverse to others.

A number of important questions arise in each of these cases: what are the
experimental bounds on the size of such extra--dimensions? do these bounds allow
the probe of extra--dimensions at future collider experiments? how can one
distinguish the corresponding signals from other possible origin of new physics,
such as models with new gauge bosons?

These questions can be addressed only in a well defined theoretical
framework, where scenaria with large extra-dimensions can be realized
consistently, such as perturbative type I string theory with
fundamental scale $M_s$ of the order of a few TeV.\footnote{For
notational simplicity, in this work we refer to type I any type
I$^\prime$ vacuum obtained by T-dualities from the original type I
theory.} Our analysis concerns then longitudinal dimensions felt by
gauge interactions, associated to compactification scales $R^{-1}$
somewhat smaller than the string scale, $R^{-1}<M_s$. The reason is
that for string size dimensions, KK (Kaluza-Klein) excitations
have masses comparable to string massive modes and their effects
become indistinguishable. On the other hand, transverse dimensions
larger than the string length are associated to superheavy open string
winding modes which decouple at low energies;\footnote{This is true if
bulk fields propagate always in more than two transverse dimensions,
or if tadpoles cancel locally \cite{ab}.} they can be probed only through
gravitational interactions associated to closed string KK modes.

Existing lower bounds on the compactification scale in
the various cases of large extra--dimensions, come from indirect
effects of KK modes in low or high energy measurements and are of the order 
of 2 to 4 TeV. These leave very little hope for production of many KK
excitations on-shell at LHC and none for other ongoing machines. In
the most optimistic case, one will be able to observe just the first
excitation, and thus, additional information or high precision   will
be necessary to bring evidence for its higher dimensional origin.

In section 2 we discuss the brane picture in models of Type I strings on
orbifolds. This will allow us to illustrate different possibilities for
localizing different parts of the observed particles on different branes. We
obtain five classes of models that we study in the following sections. In
section 3, we discuss the standard scenario where all the three gauge factors
of the standard model feel the extra-dimension. New bounds on the upper value
of the compactification scale accessible to LHC are obtained. These arise from
the increase of luminosity compared to \cite{ABQ} and the careful study of
gluon and $W_{\pm}$ channels that have not received much of attention in the
past (see \cite{Nath}). In section 4, we discuss the collider signatures of
the three other cases where only part of the standard model gauge group feels
the large extra-dimension. Section 5 deals with the 
configuration where there are no longitudinal dimensions. 
We point out that in this case the effective contact interactions are
dominated by effects due to tree-level exchange of massive string oscillation
modes and the usually considered processes of virtual KK gravitons are
sub-dominant. Finally section 6 summarizes our quantitative results.

\section{Bulk and boundary states within the brane picture}

In this work we are interested in collider experiments taking place at
energies above the electroweak scale. Effects of the electroweak breaking can
therefore be neglected and one can consider brane configurations that lead to
unbroken $SU(3)_c \times SU(2)_w \times U(1)_Y$ gauge symmetry. Although no
real realistic model has emerged yet from type I vacua \cite{type1,st}, we can
discuss the main properties associated to different possible classes of such
models.

In the generic case, the 3 group factors of the standard model may arise from
different collections of coincident branes. If the space dimensionality $p$
of a  set of branes ($p$-branes) is bigger than 3, there are  $p-3$ {\it
longitudinal} compact dimensions felt by the brane gauge interactions
associated to the KK excitations with masses 
\beq
M_n^2= \frac {n^2}{R_\parallel^2}
\label{massKK}
\eeq
for integer $n$. Here, $R_\parallel$ denotes generically the radii of
longitudinal dimensions. The remaining $9-p$ {\it transverse} dimensions are 
felt only by gravity. The brane worldvolume fields have no KK excitations 
along these directions but (superheavy) winding modes with masses 
$|n|R_\perp M_s^2$, where $R_\perp$ denotes generically the radii of
transverse dimensions. The size of compact dimensions is constrained by 
present observations to be less than TeV$^{-1}$ for the longitudinal ones 
and less than millimeter for transverse. Below, we consider a particular 
dimension with a compactification scale in the TeV region, roughly an order of
magnitude smaller than the string scale $R^{-1}\sim{\cal O}(10^{-1}) M_s$. The
3 sets of branes associated with the three standard model gauge group factors
can then be either longitudinal or transverse with respect to this particular
direction.

Since the  coupling constant of the gauge group living on the longitudinal
branes is reduced by the size of the large dimension $RM_s$ compared to those
of the transverse branes, if $SU(3)$ has KK modes all three group factors must
have. Otherwise it is difficult to reconcile the suppression of the strong
coupling at the string scale with the observed reverse situation. As a result,
there are 5 distinct cases that we denote $(l,l,l)$, $(t,l,l)$, $(t,l,t)$,
$(t,t,l)$ and $(t,t,t)$, where $l$ ($t$) refers to branes longitudinal
(transverse) to the large dimension and the three positions in the brackets
correspond to the 3 gauge group factors of the standard model $SU(3)_c\times
SU(2)_w\times U(1)_Y$. In the first two cases, we will also comment on the
anisotropic possibility of having more than one dimensions larger than the
string length. We will ignore the possibility that a gauge factor is
associated  to a linear combination of $l$ and $t$ type of branes. Since the 
longitudinal component give rise to interactions suppressed by $RM_s$,
this can be of importance only if the $t$ component is suppressed by a small 
mixing angle.

Similarly, there are three kinds of matter fields. Those which live on
longitudinal or on transverse branes and those which arise as massless
modes of open strings stretched between two different types of branes
and live on the corresponding intersections. The first two types have KK and
winding modes,  respectively, along the large dimension, while the
latter have no excitations and behave as the $Z_2$ twisted (boundary)
states of heterotic strings on orbifolds.\footnote{In contrast to the
heterotic case open strings do not lead to $Z_N$ twisted matter with
$N>2$.} The boundary states couple to all KK-modes of gauge fields in
the same way.\footnote{Actually, when one is restricted to physical
states with positive KK momentum $n$ (see eq. (\ref{massKK})),  the
coupling of massive excitations is enhanced by a factor $\sqrt{2}$
relative to the lowest (massless) one.} These couplings violate
obviously momentum conservation in the compact direction and make all
massive KK excitations unstable. If all quarks and leptons are of
the first two kinds of matter fields living on branes, their
interactions preserve the internal momentum  and KK excitations (or windings) 
can be produced only in pairs and are stable. Naively, such a possibility is
excluded, for instance  from the non--observation of stable
excitations bounded to atomes. However,  this  assumes their
production during the  early history of the universe which might be
suppressed if the  ``maximal'' temperature is very low as suggested in
\cite{lowT}. In the following we ignore this possibility and assume
that all matter fields are localized in the large dimension, with the
exception of the $(t,t,t)$ case which we treat in section 5.

In addition to the KK excitations of massless modes there might be massive
modes with different quantum numbers. Their corresponding massless modes are
projected out since they transform non-trivially under the  orientifold
projection that leads to chiral spectrum. These {\it odd} states can be new
gauge bosons that enlarge the gauge symmetry or extra fermions and scalars.
They can decay only to  localized matter fields, but their coupling vanishes
to lowest order. Non-vanishing couplings can be obtained by taking derivatives
along the  extra-dimension and are therefore suppressed by powers of $RM_s$.
For instance, the lowest order of such operators (in non-supersymmetric
models) involves three scalars one of which is odd and has a derivative
acting  on it. These states are in general model dependent and will not be
discussed here.

\section{ The ``standard'' $(l,l,l)$ scenario: }
 
In the $(l,l,l)$ scenario all gauge bosons of the standard 
model $SU(3)_c\times SU(2)_w\times U(1)_Y$ have KK excitations.
Because matter fields are localized, their interactions do not preserve the 
momenta in the extra-dimension and single KK excitations can be produced.
This means for example that QCD processes $q{\bar q} \rightarrow G^{(n)}$
with $q$ representing quarks and $G^{(n)}$ massive KK excitations 
of gluons are allowed. In contrast processes such as $G G \rightarrow G^{(n)}$
are forbidden as gauge boson interactions conserve the internal momenta.
The exchange of virtual KK modes leads to effects in low energy processes
that are constrained by fits to high precision data \cite{scc,Delgado}. For
instance,  fit of measured values of $M_W$, $\Gamma_{ll}$ and $\Gamma_{had}$
lead to $R^{-1} \simgt 3.5$ TeV. Inclusion of $Q_W$ measurement, which does
not give a good agreement with the  standard model itself, raises the bound to
$R^{-1} \simgt 3.9$ TeV \cite{Delgado}.

For the LHC experiment, this scenario predicts that new
resonances in three different channels $l^+ l^-$, $l^{\pm} \nu$ and dijets
could be observed at the same mass if the compactification scale is low enough.
The resonances  due to the $SU(3)$ gluon excitations  have quite large widths 
due to the ``strong'' coupling value. They are thus spread  and  difficult to
detect  already for compactification scales of the order of 5 TeV. In figure 1
we show the shape of a typical expected signal for the production of a gluon
excitation for $R^{-1} = 4$ and 5 TeV at the LHC  (assuming $\sqrt{s}=14$ TeV
and $L=100 fb^{-1}$). Our results  have been obtained using the CTEQ4L
structure funcions \cite{ctq4l}. We have also included the multiplicative
K-factor \cite{Kfactor}, taken to be $K=1.1$, and summed over all jets,
top excluded. Moreover, we have implemented a rapidity cut, $|\eta |\le$ 0.5,
on both jets and required the invariant mass to be $M_{jj'}\ge$ 2 TeV.

The strength of the coupling constant of gluon excitations implies on the 
other hand a large excess in the total number of events from the QCD expected
ones. We found that looking for excess of dijet events  could be the most
efficient channel to constrain the size of extra-dimensions. In figure 2, we
present the effect of these gluonic KK states on the total number of dijet
events. The significance, given by the ratio $|N_T-N_{SM}|/\sqrt{N_{SM}}$
where $N_T$ is the total number of events and $N_{SM}$ is the SM background,
indicates that LHC could probe in this hadronic channel values of the
compactification scale up to about 20 TeV. This becomes of order 15 TeV for a
lower luminosity of $L=10 fb^{-1}$ used in \cite{ABQ}. The Tevatron run II 
with luminosity of $L=20 fb^{-1}$ will probe compactification scales of 
order 4 TeV.

The KK-excitations $W_{\pm}^{(n)}$ of the W bosons could show
up in clean final states consisting of one charged lepton and missing
energy ($\nu_l l$ with $l=e,\mu$). The differential cross-section in the
transverse mass has the shape illustrated for $R^{-1} = 4, 5$ and 6 TeV in 
figure 3. We have required the charged lepton to be in the central region,
$|\eta|\le$ 1. This cut reduces the SM background by a factor of about 50$\%$,
while decreasing the pure signal only by $35\%$ due to the large masses
considered. The discovery limit of such a resonance is  around 6 TeV. 

Approximately the same values of compactification scales $\simlt 6$ TeV 
could be discovered through the observation of a resonance in the $l^+ l^-$ 
final state (with $l = e, \mu$). The interference in the overlap of the 
resonances $\gamma^{(n)}$ of the photon and $Z^{(n)}$ leads to an
 a priori distinctive deep as noticed in \cite{ABQ,Nath}. 
The shape of the resonance is quite different from the one of a Bright-Wigner
due to the excitation of only one gauge boson, as can be seen in figure 3.  
 
In a way similar to the case of excitations of gluons, the exchange of
 virtual KK excitations  $W_{\pm}^{(n)}$, $\gamma^{(n)}$ and $Z^{(n)}$
 lead to deviations from the standard model expectation in the number
 of total events. These allow to probe higher  compactification
 scales. Figure 4 shows that at $95\%$ confidence level LHC could
 exclude values of compactification scales up to 12, 14 and 15 TeV
 from the photon+$Z$, $W_{\pm}$ and combined channels, respectively.
Here, for the dilepton final state, we have required one lepton to be in  
the central region, $|\eta_l|\le$ 1, the other one having a looser
cut $|\eta_{l^\prime}|\le$ 2.4. This gives about the same acceptance
as above. Moreover, we have chosen a 400 GeV lower bound on the transverse
and invariant mass, in order to optimize the significance.

\section{The cases $(t,l,l)$, $(t,l,t)$ and $(t,t,l)$: }

The difference between the $(t,l,l)$ and the $(l,l,l)$ scenario is that 
in the former the gluons do not feel the extra-dimensions. Only
the KK excitations  $W_{\pm}^{(n)}$, $\gamma^{(n)}$ and $Z^{(n)}$ are present 
and lead to the same effects as for the $(l,l,l)$ case. The limits
on the extra-dimensions follow from the previous section: 6 TeV for discovery 
and 15 TeV for the exclusion bounds.

In the $(t,l,t)$ case, only the $SU(2)$ factor arises from a set of
branes longitudinal to the TeV-scale extra-dimension. As all the states 
charged under $U(1)_Y$ are localized in this case, one 
has no more the freedom to choose the Higgs to be  localized or not as 
it was for the cases $(l,l,l)$ and $(t,l,l)$ but only the first possibility 
exists.  
The massive
modes are  KK excitations of $W_{\pm}$ and $W_3$. The latter will lead
to a deficit  in $l^+ l^-$ half of the corresponding one in the $l^{\pm}
\nu$ channel (as two gauge bosons contribute here). Moreover, the 
forward-backward asymmetry is maximal. The limits on this
scenario arise from exchange of  $W_{\pm}$ and are again 6 TeV for
discovery and 14 TeV for the exclusion  bounds.

The third case is the one of $(t,t,l)$ channel where only $U(1)_Y$ feels 
the extra-dimension. Similarly to the previous case, the Higgs doublets have 
to be identified with localized states. In this case the limits are weaker, the
exclusion bound is in fact around 8 TeV, as can be seen in
figure 6. 

Finally, let us comment about some of the possibilities that we
neglected. In the scenaria discussed above, we have assumed that only
one  extra-dimension is large and longitudinal to some sets of
branes. In the $(t,l,l)$ and the $(l,l,l)$ cases one could  
imagine that the different gauge factors arise from different sets of branes
feeling different dimensions. In this case, the $\gamma^{(n)} +Z^{(n)}$
resonances split into resonances of
$W^3$ and $U(1)_Y$ located at different masses. Another possibility to
notice is that the case where part of $U(1)_Y$ is  $t$ and part is
$l$, one would observe a signal which is hard to distinguish from a
generic extra $U(1)'$. A good statistic would be needed to distinguish
the deviation in the tail of the resonance as being due to effects
additional to those of the $U(1)'$ itself.

\section{Limit on the string scale:}

Finally we discuss the case $(t,t,t)$ where all the extra-dimensions are
transverse to the 3-brane where the standard model lives.  
In this case standard model particles have no KK excitations and the 
main experimental signals are due either to gravitational effects 
associated to  closed strings propagating in the ten-dimensional bulk or to 
the exchange of massive string oscillation modes. The gravitational effects 
were extensively studied in the litterature based on graviton exchange 
in the effective field theory \cite{lowgrav,lykken2,Hewett}. In the context
of Type I string models, these effects are part of the one-loop diagram of
open strings (annulus) that  can be  seen as closed strings exchanged in the
transverse channel (cylinder). However these are subdominant, ${\cal O}
(g_s^2)$ with $g_s$ the string  coupling, compared to the {\it tree level}
exchange of  open string oscillation modes, which are of order ${\cal O} (g_s)$
\footnote{This observation is also made recently by M. Peskin in \cite{Peskin}.}. 

Indeed, the  exchange of virtual gravitons is described in the effective 
field theory by an amplitude of the form ${\cal S}(s) {\cal T}$ with 
\cite{lowgrav}:
\beq
{\cal T} = T_{\mu \nu}T^{\mu \nu} - \frac {1}{1+d_\perp} T_\mu^\mu T_\nu^\nu
\label{enertens}
\eeq 
and 
\beq 
{\cal S} = \frac {1}{M_p^2} \sum_n \frac
{1}{s-\frac{{\vec n}^2}{R_\perp^2}} 
\eeq 
where $T^{\mu \nu}$ is the
energy momentum tensor and $s$ the center of mass energy. The sum in ${\cal S}$ is  divergent for a
number $d_\perp > 1$ of transverse dimensions. From
the string theory point of view this corresponds to an ultraviolet 
one-loop divergence
cut-off appropriately by the string scale. The final  result 
is finite and model (compactification) dependent. A phenomenological approach
followed in \cite{IK,lykken2,Hewett} is to estimate ${\cal S}$ as 
\beq 
{\cal S} = g_s^2 \frac {A} {M_s^4}
\label{QFT}
\eeq 
where $A = \log{ \frac{M_s^2}{s}}$ for $d_\perp =2$ and 
$A = \frac{2}{d_\perp-2}$ for $d_\perp > 2$. The coupling constant 
factor $g_s^2$ arises from the Type I relation $M_p^2= M_s^2 
(M_s R_\perp)^{d_\perp}/g_s^2$ and the fact that the 
divergent  sum is cut-off at the string scale $M_s$.

On the other hand, the tree-level open string four-point amplitude is:
\beq
{\cal A}(1,2,3,4)= \frac{1}{ M_s^2}g_s 
\frac{\Gamma(1-s/M_s^2)\Gamma(1-t/M_s^2)}{\Gamma(1- s/M_s^2 -t/M_s^2 )} 
K(1,2,3,4) + [ s \rightarrow t] + [ s \rightarrow u ]
\label{treest}
\eeq
where $K$ contains  kinematic factors as well as gauge indices (Chan-Paton 
factors) \cite{Polch}. We have introduced the Mandelstam kinematic invariants:

\beq
s= -(p_1+p_2)^2 \, \, \, \, \, \, t= -(p_1+p_3)^2 \, \, \, \, \, \, 
u=-(p_1+p_4)^2
\label{Mand}
\eeq
with $p_i$ the momentum of state $i$. 
Expanding the Gamma-functions at low energies 
in powers of $s/M_s^2$, and summing over the three permutations, one finds
for the 4-fermion amplitude:
\beq
{\cal A}(1,2,3,4)\simeq
g_s\frac {{\cal A}_s}{s} (1 +\frac {\pi^2}{12} \frac {s^2}{M_s^4}) +
g_s\frac {{\cal A}_t}{t} (1 +\frac {\pi^2}{12} \frac {t^2}{M_s^4}) +
g_s\frac {{\cal A}_u}{u} (1 +\frac {\pi^2}{12} \frac {u^2}{M_s^4})\ .
\label{form}
\eeq
Note that the leading contribution
$g_s(\frac{{\cal A}_s}{s}+\frac{{\cal A}_t}{t}+\frac{{\cal A}_u}{u})$
reproduces the point-like particle result. In ${{\cal A}_s}$ we have absorbed
the kinematical as well as the trace on Chan-Paton trace factors:
\beq
{{\cal A}_s} = Tr (\lambda^1 \lambda^2  \lambda^3 \lambda^4)
{\bar u_1}\gamma^\mu u_2 {\bar u_3}\gamma_\mu u_4
\eeq
where $\lambda^i$ are the Chan-Paton factors, $u_i$ are the associated spinors 
and we dropped numerical factors. The ${{\cal A}_t}$ and ${{\cal A}_u}$
 are obtained by replacing the cyclic order $1234$ by $1342$ and $1423$
respectively. For example, in the usual Bha-Bha 
scattering, 
${{\cal A}_u}$ vanishes due to the traces on Chan-Paton factors and 
the two terms with poles in  $s$ and $t$ channels remain. The next terms 
describe
effective contact interactions. A comparison with eq.(\ref{QFT}) shows
that they are enhanced by a string-loop factor $g_s^{-1}$ with respect to
the field theory estimate for KK graviton exchanges.
Although the precise value of $g_s$ requires a detailed analysis of threshold 
corrections, for a rough estimate one could take $g_s\simeq\alpha\sim 1/25$, 
that implies an enhancement by an order of magnitude. 

As a result, one can not get reliable experimental  bounds on the string scale
just based on fits of differential cross-sections with field theoretical
estimates for the  contribution of graviton KK exchange, but should use
processes with missing energy due to emission of ``on-shell''  gravitons in the
bulk \cite{lowgrav}. Alternatively, the four-point contact interactions
analyzed above may also be used to provide  direct bounds on the string scale.
Note however that these interactions are in general model dependent.

Above we have considered the case of 4-fermion
interactions which receive lowest order contributions at tree-level from
massless fields. However, one could also consider processes for
which the dominant lowest order contributions arise from exchange of massive 
open string oscillator modes. Such an example is the four-point amplitude
of abelian gauge fields, which gives rise to leading contact interactions
encoded in the Born-Infeld action \cite{Polch}.

\section{Conclusions}

We have described various possible models allowed by the
Type I picture of the brane-world scenario. Assuming that
matter fermions are localized in some large TeV extra-dimension and that
there is no inverse hierarchy of the tree-level gauge couplings compared to
the observed (loop-corrected) ones, we found 5 classes of models,
denoted as $(l,l,l)$, $(t,l,l)$, $(t,l,t)$, $(t,t,l)$ and $(t,t,t)$.
For these models we have attempted to answer the questions raised in the
introduction. What are the experimental bounds on the size of such
extra--dimensions? Lower bounds on the compactification scale from low energy
measurements are of the order of 3 to 4 TeV. 

Do these bounds allow the probe of extra-dimensions at future collider
experiments?  We have found that new resonances can be discovered at LHC for
scales as high as around  6 TeV for all the four cases $(l,l,l)$, $(t,l,l)$,
$(t,l,t)$ and $(t,t,l)$. This implies that only a small window of 3 - 4 to 6
TeV for the compactification scale is left. Improvement of low energy
precision bounds from existing experiments (LEP, Tevatron...) are thus of
great importance as they may allow to narrow further this window. Higher
scales can be probed through indirect effects of exchanges of virtual KK
excitations. We found that the most efficient channel 
is through KK excitations of gluons which give sensitivity up to 
$R^{-1}\simlt 20$ TeV. The  $W_{\pm}$, photon+$Z$  and  $U(1)_Y$ boson allow 
to probe values $R^{-1}\simlt 14$, 12 and 8 TeV, respectively.

How can one distinguish the corresponding signals from other possible origin
of new physics, such as models with new gauge bosons? The discussion
on this issue follows the one for the $(t,l,l)$ case in \cite{ABQ,Nath}.
We have pointed out that scenaria $(l,l,l)$, $(t,l,l)$ and $(t,l,t)$
predict the existence of resonances in three or two final channels.
As for the $(t,l,l)$ case in \cite{ABQ,Nath}, we have pointed out that 
models $(l,l,l)$ and $(t,l,t)$ predict the existence of resonances in three
and two final channels, respectively. They are located at the same value of
energy. This property is not shared by most of other new gauge boson models.
Moreover, the heights and widths of the resonances are directly related to
those of standard model gauge bosons in the corresponding channels. Also,
in the case of excitations of the photon+$Z$ a deep due to the interference
between the two bosons should be observed just before the resonance.

For the last case $(t,t,t)$, we pointed out that there are open-string
tree-level contact terms that are dominant with respect of those induced by
the exchange of KK excitations of gravitons. Analysis of these terms
should allow in principle to get stronger bounds on the string scale. The
precise coefficients, however, of the higher-dimensional effective operators
are in general model (compactification) dependent and further investigation
is needed.

\vskip0.5cm

We wish to thank T. Kamon and Y. Oz for discussions at various stages of the 
work. KB wishes also to thank
A. Delgado, A. Pomarol, M. Quiros and J. Lykken  for discussion. 
This work is supported in part by the EEC under TMR contract 
ERBFMRX-CT96-0090 and in part by NSF grant No. PHY-9722090.

\newpage

\begin{figure}[htb]
\centering
\epsfxsize=6.5in
\hspace*{0in}
\epsffile{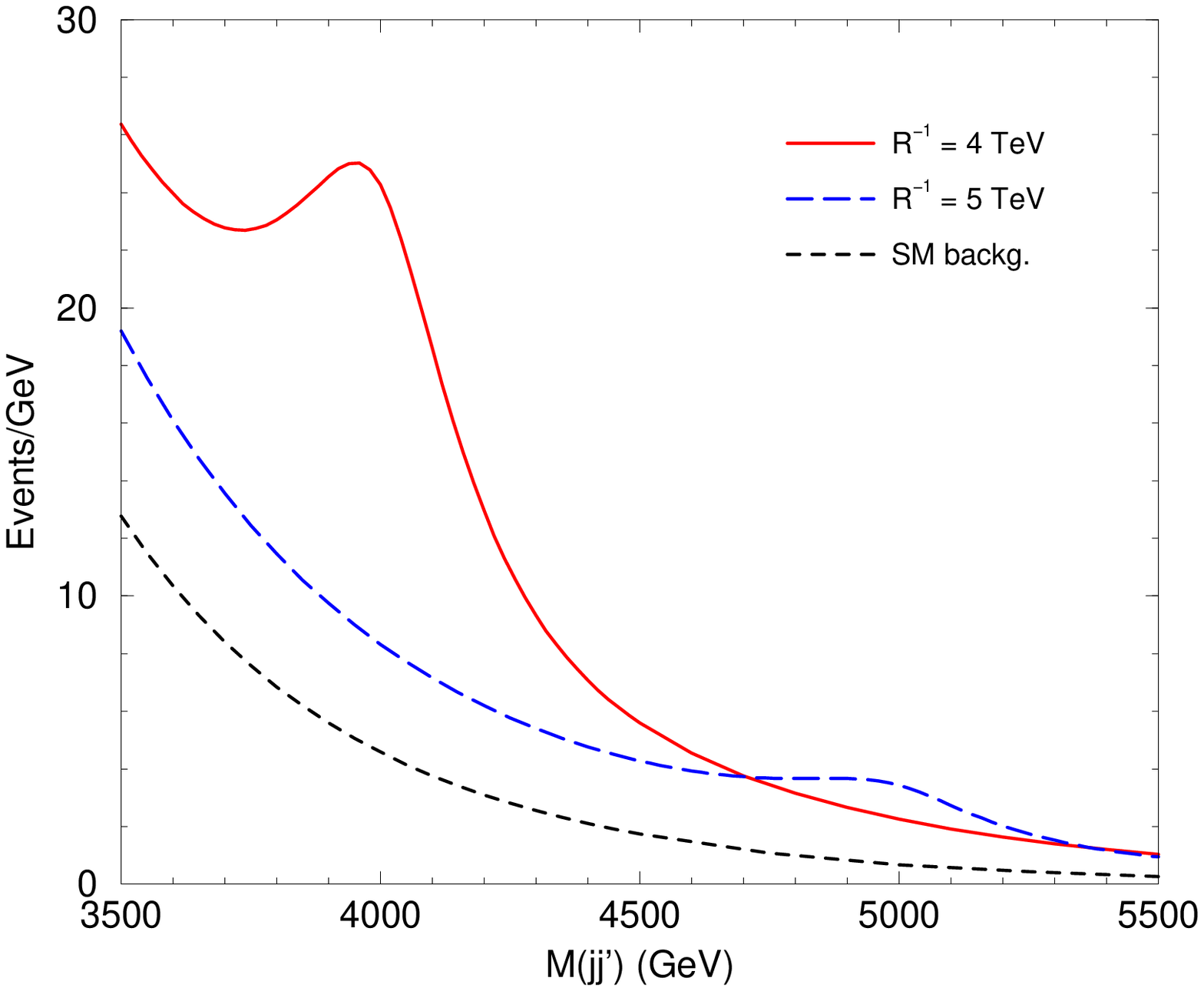}
\caption{\it First resonances in the LHC experiment due to a KK excitation 
of gluon for one extra-dimension at 4 and 5 TeV.}
\label{fig:fig1}
\end{figure}

\begin{figure}[htb]
\centering
\epsfxsize=6.5in
\hspace*{0in}
\epsffile{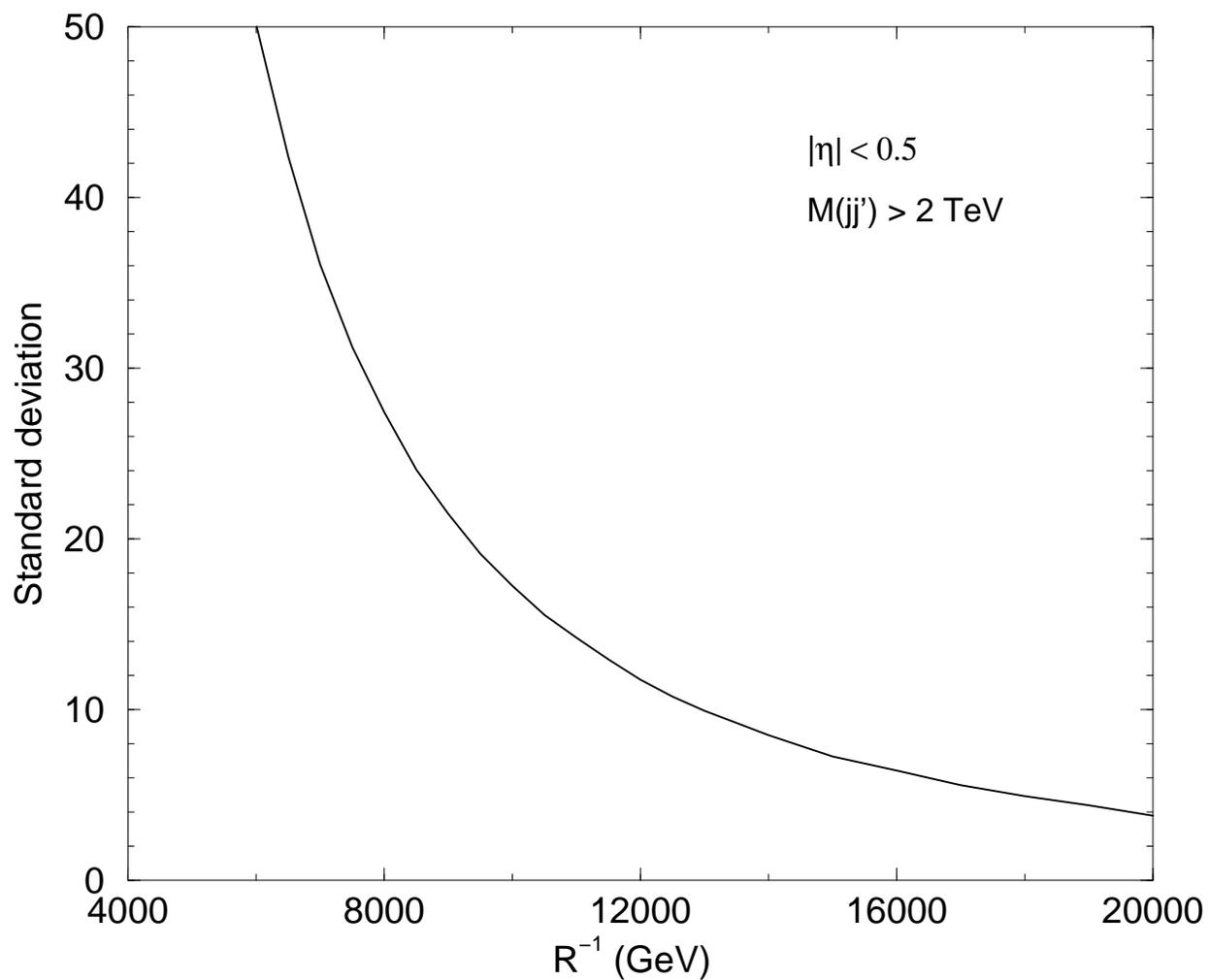}
\caption{\it Number of standard deviation in number of observerd dijets from 
the expected standard model value, due to the presence of a TeV-scale 
extra-dimension of compactification radius $R$.}
\label{fig:fig2}
\end{figure}

\begin{figure}[htb]
\centering
\epsfxsize=6.5in
\hspace*{0in}
\epsffile{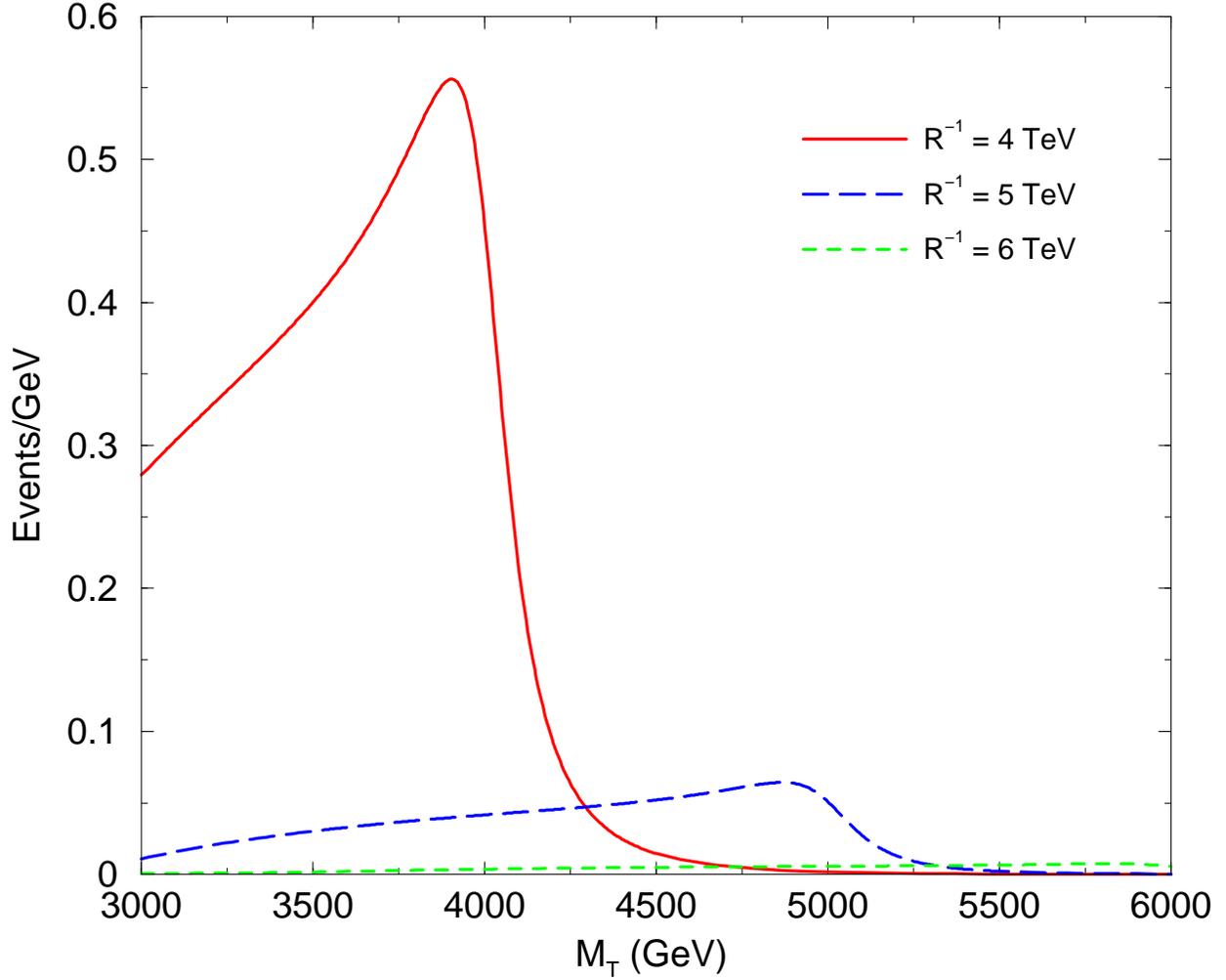}
\caption{\it First resonances in the LHC experiment due to a KK excitation 
of $W_{\pm}^{(n)}$ for one extra-dimension at 4, 5 and 6 TeV. We plot the 
differential cross section as function of the 
transverse mass for the $W$s.} 
\label{fig:fig3}
\end{figure}
\begin{figure}[htb]
\centering
\epsfxsize=6.5in
\hspace*{0in}
\epsffile{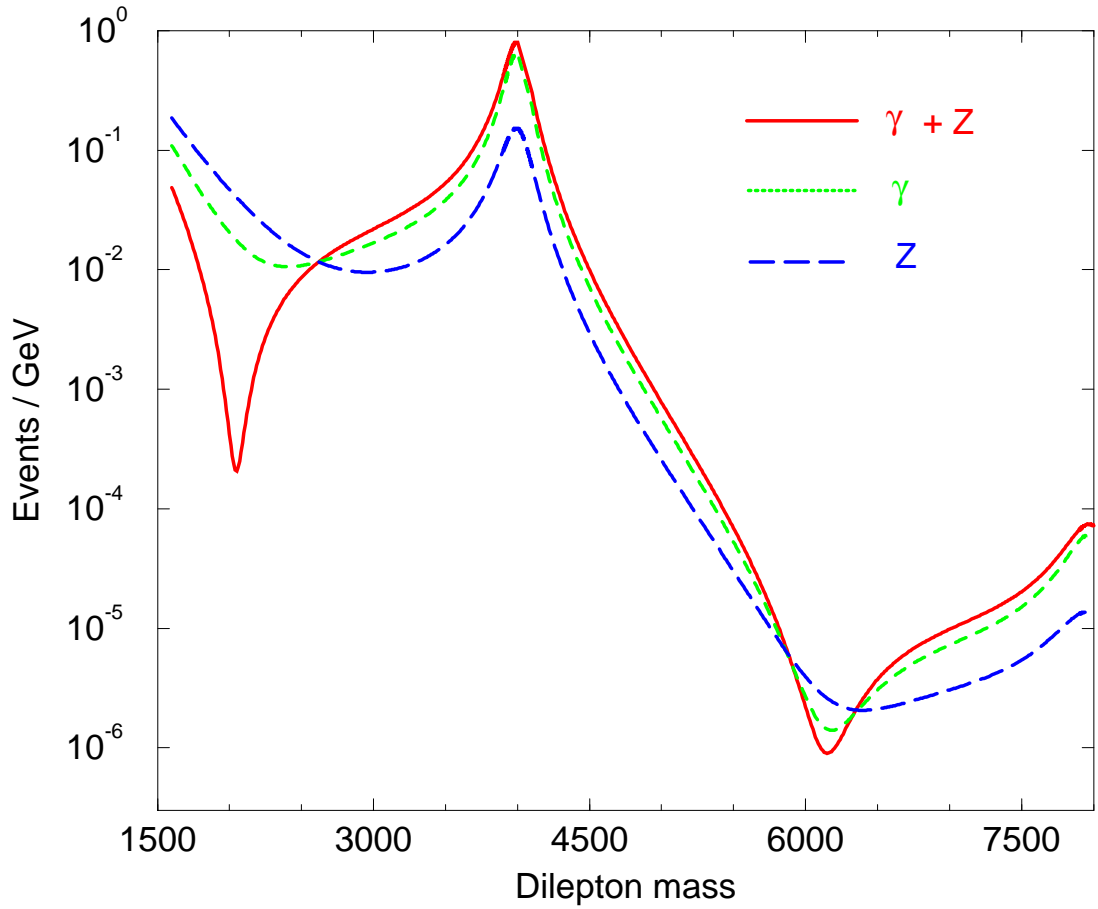}
\caption{\it First resonances in the LHC experiment due to a KK excitation 
of photon and Z  for one extra-dimension at 4 TeV. From highest to lowest: 
excitation of photon+Z, photon and Z boson.}
\label{fig:fig4}
\end{figure}

\begin{figure}[htb]
\centering
\epsfxsize=6.5in
\hspace*{0in}
\epsffile{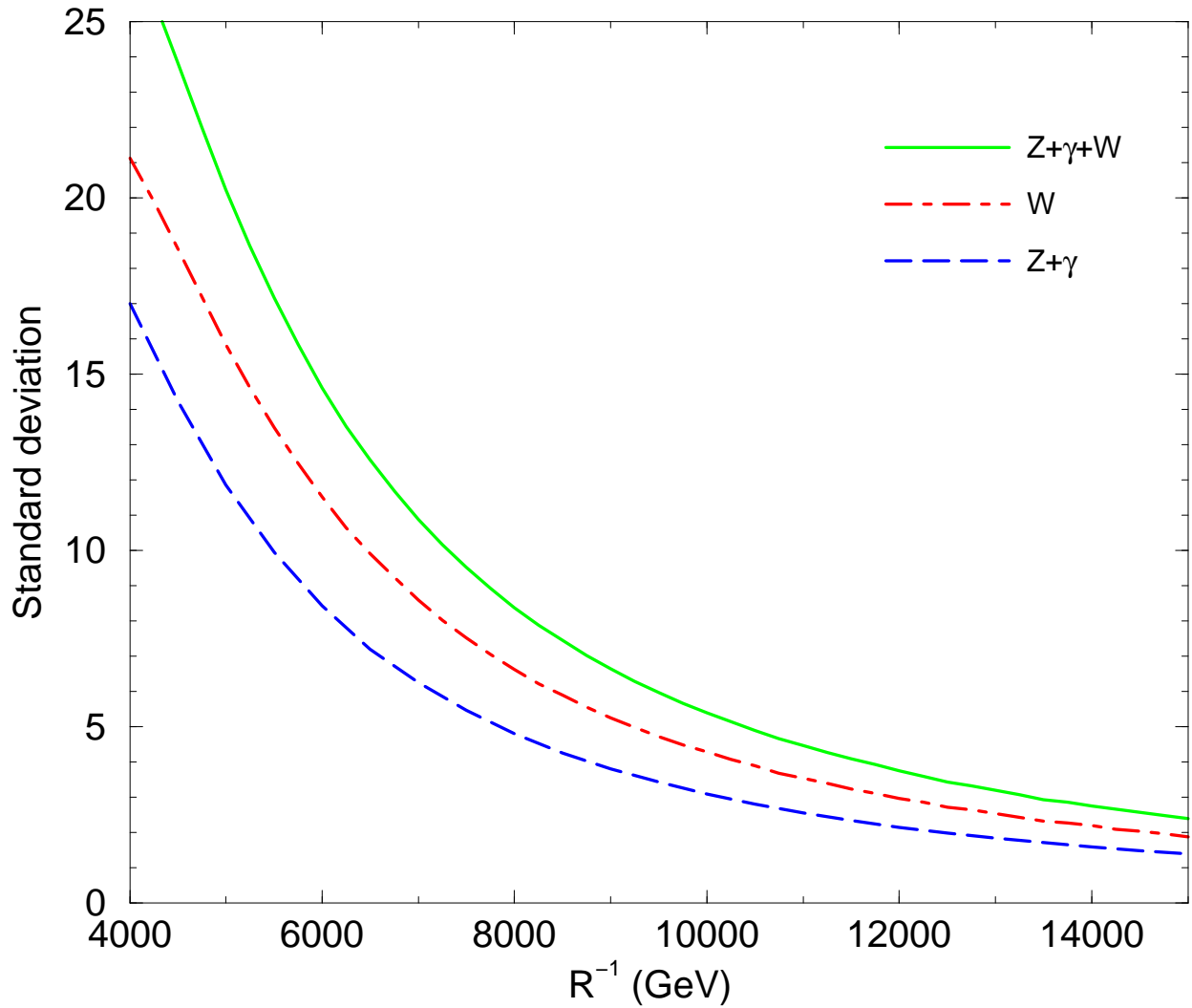}
\caption{\it Number of standard deviation in the number of  $l^+ l^-$ pairs  
and $\nu_l l$ pairs produced from the expected standard model value due to the presence of one extra-dimension of radius $R$.  }
\label{fig:fig5}
\end{figure}

\begin{figure}[htb]
\centering
\epsfxsize=6.5in
\hspace*{0in}
\epsffile{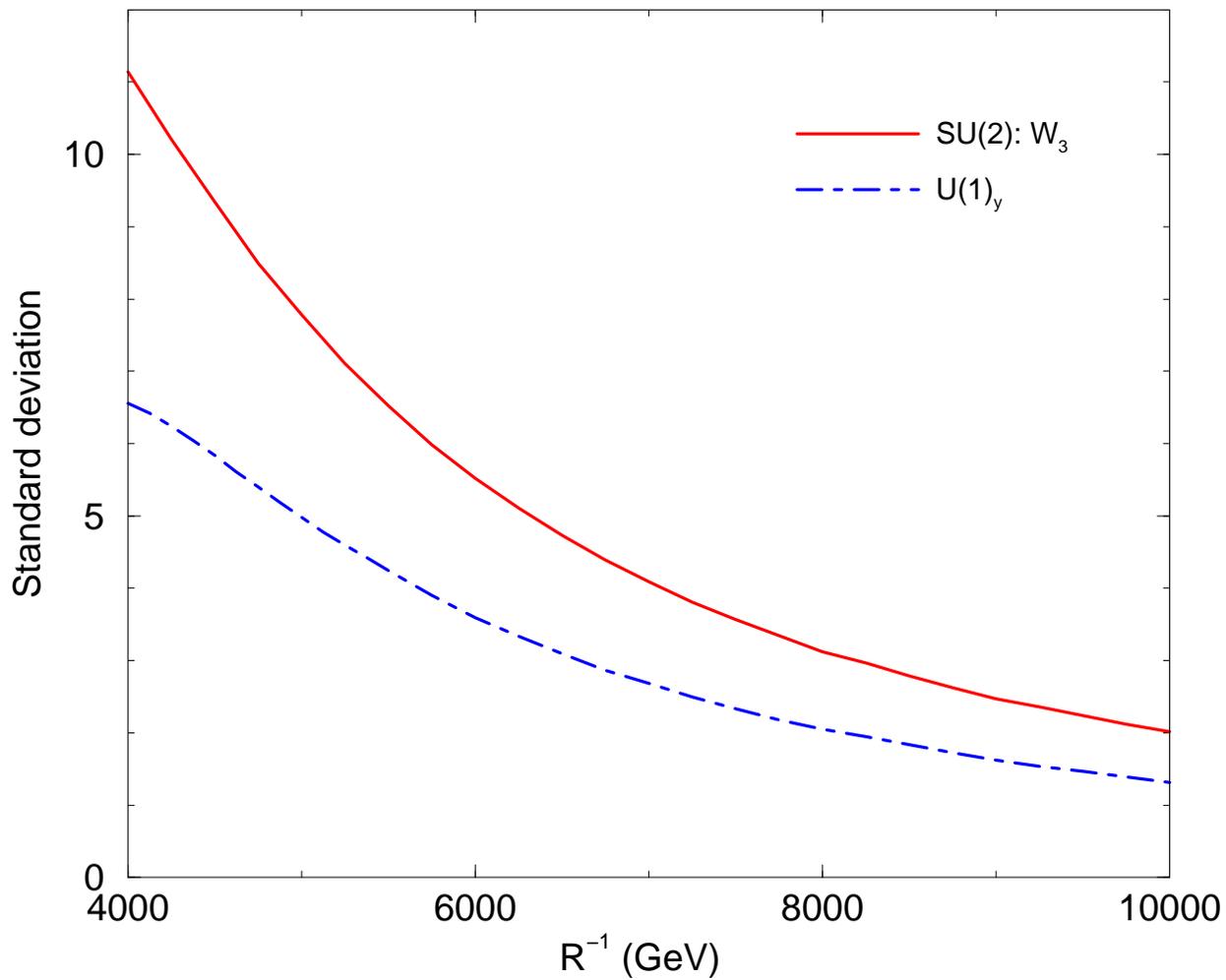}
\caption{\it Number of standard deviation in the number of  $l^+ l^-$ pairs  
produced from the expected standard model value due to the presence of 
one extra-dimension of radius $R$ in the case of $(t,l,t)$ i.e. $W_3$ 
KK excitations and the case
of $(t,t,l)$ i.e. $U(1)_Y$ boson KK excitations.  }
\label{fig:fig6}
\end{figure}

\end{document}